Strong angular dependent transmission ratio due to interplay between spin waves and the domain wall in perpendicular magnetic anisotropy materials

Liang-Juan Chang[1], Yen-Fu Liu[1], Ming-Yi Kao[1], Li-Zai Tsai[1], Jun-Zhi Liang[2, a)], and Shang-Fan Lee[1, b)]

[1] Institute of Physics, Academia Sinica, Taipei 11529, Taiwan
[2] Department of Physics, Fu Jen Catholic University, Taipei 24205, Taiwan

**Abstract**

Spin waves (SW) can induce rich domain wall (DW) motion in a perpendicular-magnetic anisotropy nanostrip. In-plane magnetization tilt angle resulting from the fluctuation of the effective field of the magnetization response in the DW region plays an essential role in the dynamics of SW interacting with a DW. We performed simulation and found that the transmission ratio of the propagating SW across the DW depends strongly on the tilt angle in the low-frequency regime. The material parameters and the geometrical configuration can be fine-tuned for practical devices.

a) Email: phys2021@mail.fju.edu.tw
b) Email: leesf@gate.sinica.edu.tw

Manipulation of magnetic domain walls (DW) in ferromagnetic nanostrips has been intensively investigated in view of both fundamental research and potential for technological applications. Several DW based devices, including spintronic logic [1] and magnetic memory devices [2, 3], have been proposed. Moving the DW in a controlled manner is an important issue in those applications. Employing external magnetic fields via energy dissipation [4, 5] and spin polarized electric current via momentum transfer [6-8] are well known means for DW motion. Recently, to overcome the issue of Joule heating in current-driven magnetization reversal, many research groups have proposed that the propagating spin waves (magnons) in a ferromagnetic nanowire are able to assist magnetic domain wall motion [9-14]. Spin waves (SW) can drive the DW effectively since they consist of magnonic spin current. It has been theoretically shown that magnonic spin-transfer torque (STT) causes a DW to propagate in the direction opposite to SWs [10] and that the linear momentum transfer from magnons causes a DW to propagate in the direction of SWs [12]. The former occurs in one-dimensional (1D) systems when SWs have to transmit through a DW [10] and the latter when SWs are reflected by a DW in 2D nanostrips [11-14]. Conversely, manipulation of SWs also attracted much attention [15-24]. We show here by simulation that the dispersion of DW motion is much more complex due to the rotation of the magnetization inside the DW, and the transmission ratio of the SW amplitude depends on the DW orientation in perpendicular-magnetic anisotropy (PMA) nanostrips.

Magnons can be considered as spin-1 bosons with angular momentum $\pm\hbar$ and linear momentum $\hbar\mathbf{k}$ [10, 25]. When the SW passes through DWs, the magnonic spin current changes its sign. As a result, there is a spin angular momentum transfer from the propagating magnons to the DW, which generates a torque and induces the DW motion opposite to the SW to absorb this torque. When the SW is partially or completely reflected [12, 13], the linear momentum transfer of the SW reflected at the magnetic

DW induces an effective field which results in the rotation of the DW magnetization plane and forward motion of the DW [13, 26]. Interactions between the SW and the DW are related to the geometry and dimensions of magnetic elements and their material parameters such as the magnetic anisotropy constant and the saturation magnetization. However, in contrast to the cases of polarized charge current-induced domain wall motion, the SW-driven DW motion is still under development [25].

In a micromagnetic simulations study by Wang *et al.* [12], the dependence on DW width and SW frequency (wavelength) of SW propagation and DW motion were investigated in a PMA strip with a Bloch DW. They found, for large anisotropy or a narrow DW, the complete transmission of SWs in the high-frequency range. For small anisotropy or a wide DW, the complete transmission of SWs was extended to lower frequencies, even close to the cutoff frequency. They proposed that a dynamic stray field, which arises when a SW travels in a DW causing surface magnetic charges to appear on the DW boundaries, was the reason for the reflection of SWs. This field was approximated using the demagnetization factor $N_{DW}^{X}$, which was determined by the DW width, $\Delta = \pi\sqrt{A/K_{\perp}}$, where A is the exchange stiffness, and $K_{\perp}$ is the perpendicular anisotropy constant [13]. The reflectivity of SWs was manipulated by changing the anisotropy constant and hence the DW width. Although the dependence of DW motion on DW width and SW frequency has been thoroughly investigated, the DW orientation has not been considered at length. It is necessary to release the constraint of rigid DWs and study how SW propagation and DW motion depend on the DW's internal structures.

In this work, we study how the propagation of the SW changes the orientation of the DW's magnetization inside a Néel wall and the DW motion, as well as the variation of the transmission ratio of the SW due to the DW dynamics in PMA nanostrips. When the SW wavelength is larger than the DW size, the Walker breakdown [4] and the transmission ratio of the SW passing through the wall are strongly dependent on the orientation of the DW. We also calculate the transmission ratio of the SW due to the DW dynamics by a simple 1D model that fit the simulation results qualitatively well.

The PMA materials we present here are CoFeB [27, 28] and NiFe [29]. The nanostrip studied is 4 μm long in the x direction, as shown in Fig. 1. For CoFeB, the strip width w is 50 nm and thickness t is 1 nm, following the experimental Ta(5 nm)/$Co_{20}Fe_{60}B_{20}$(1 nm)/MgO(2 nm)/Ta(5 nm) structure. For NiFe strip, w is 30 nm and t is 5 nm. To study the SW and DW dynamics, micromagnetic simulations have been performed with the micromagnetic code OOMMF [30] with a unit cell of $2 \times 2 \times 1$ for CoFeB and $2 \times 2 \times 5$ nm$^3$ for NiFe. The values of material parameters for CoFeB and NiFe were respectively saturation moments $M_S$ = 8.75 and $8.6 \times 10^5$ A/m, exchange stiffness constants $A$ = 1.0 and $1.3 \times 10^{-11}$ J/m, the perpendicular anisotropy constants $K_\perp$ = 5.1 and $5.8 \times 10^5$ J/m$^3$, and the damping parameter $\alpha$ = 0.01 for both. A 180° DW was first introduced into the nanostrip by applying a sequence of alternating magnetic fields along the magnetic easy axis (z) and then returning to the remnant state. The resulting DW is a Néel wall. The DW at the center of the strip is subjected to a SW source 0.5 μm apart on the left. The SWs are excited locally in an area 2 nm across, shown as the green part in Fig. (1), by a harmonic sinusoidal field $H = H_0 \sin(2\pi f t) e_y$ with amplitude $H_0$ in the transverse direction $y$ and frequency $f$. There is no dc external magnetic field when the SW is active. The nanostrip serves as a waveguide for the SW with a cutoff frequency determined by the dispersion relation [25]. The cutoff frequency $f_c$ is around 3.2 GHz for CoFeB and 14.3 GHz for NiFe in our cases.

The time evolution of the DW motion driven by SWs was obtained as shown in Fig. 2. Based on the characteristic profiles of the DW displacement versus time curves, the DW motion induced by the SW can be separated roughly into three regions with increasing SW amplitude. They are forward, oscillatory, and backward motions and combinations on the borders as shown in Fig. 2 (a) for CoFeB with $f = 5$ GHz and in Fig. (b) to (e) for NiFe with $f = 20$ GHz. The initial transient backward motion for all cases is associated with the separation between the DW and the SW source. We plot the profile of dynamic DW motion for NiFe with the displacement x as a two-dimensional phase diagram of $H_0$ and $t$, as shown in Fig. 2(f). The CoFeB shows similar phase diagram except near the boundaries. As we describe in the following paragraphs, when the SW is reflected [12, 13], an effective field is induced resulting in rotation in addition to the motion of the DW. The transmission ratio of SW is dependent on the DW orientation that in turn influence the DW motion. By these interactions between SW and DW, the domain motion could become very complex.

For region I, forward displacement after the initial transient motion was obtained in the wide range with small excitation amplitude shown as black line for $H_0 = 120$ mT, red line for $H_0 = 130$ mT, and blue line for 135 mT in Fig. 2(a) for CoFeB and in Fig. 2(b) for $H_0 = 200$ mT for NiFe. These motions are associated with changes of azimuthal angle, $\delta\varphi$, in the DW structure, as discussed in the supplemental material [31]. CoFeB shows region I behavior when $H_0 \leq 138$ mT. The damped oscillatory motion of the DW is due to the relativity larger attenuation length of the propagation SWs in this material. Consequently, the disturbances of the SWs have long-distance influence on the DW. NiFe shows region I behavior when $H_0 \leq 270$ mT. The $\delta\varphi$ was observed to be less then 45° in this region.

Region II is very narrow in the phase space, 270 mT $\leq H_0 <$ 288.8 mT for NiFe, where the DW shows backwards motion though the final displacement is forward, as shown in Fig. 2(c) for $H_0 = 280$ mT. . This phenomenon is related to the transformations between different types of DWs for the cases of $H_0 > H_w = 270$ mT, where $H_w$ is the Walker breakdown field. The DW plane rotations cause the demagnetizing torque and instantaneous velocity to change direction. The largest $\delta\varphi$ was between 45° and 270° in this region. Fig. 2(d) shows the localized steady state oscillatory motion of the DW for $H_0 = 288.8$ mT, corresponding to the boundary of region II and region III. The amplitude of this oscillatory motion is 140 nm and the period is 32 ns. The $\delta\varphi$ rotates 360° as shown in Fig. 3. CoFeB shows region II behavior when 139 mT $\leq H_0 \leq$ 139.9 mT. We did not find localized oscillatory motion with 0.01 mT resolution between regions II and III.

In region III, the shapes of x versus t curve show oscillatory motions of the DW associated with propagations in the opposite direction to the SW as shown by the pink line for $H_0 =$139.93 mT, green line for $H_0 =$140 mT, and purple line for $H_0 =$160 mT in Fig. 2(a) for CoFeB and in Fig. 2(e) for NiFe with $H_0 = 320$ mT. The DW acquires a negative average velocity, moves backwards, and is finally trapped at the SW source. In this situation, magnonic STT plays a crucial role in SW-induced DW motion. For larger $H_0$, the structure of the DW is destroyed by the larger amplitude of the SW and the rigid DW approximation is no longer valid. The total demagnetization torque vanishes due to the irregular structure of the DW.

The rotation of the DW plane plays a crucial role in the DW dynamics. The dynamics of the local magnetization when the SW is present is described by the modified Landau-Lifshitz-Gilbert (LLG) equation [12],

$$\frac{\partial \mathbf{M}}{\partial t} = -\gamma \mathbf{M} \times \mathbf{H}_{\text{eff}} + \frac{\alpha}{M_s} \mathbf{M} \times \frac{\partial \mathbf{M}}{\partial t} - \frac{\partial \mathbf{J}_m}{\partial x}, \qquad (1)$$

where $\gamma$ is the gyromagnetic ratio, $\alpha$ is the Gilbert damping parameter, $M_s$ is the saturation magnetization, $\mathbf{H}_{\text{eff}}$ is the effective magnetic field consisting of anisotropy, demagnetization, and exchange fields, and $\mathbf{J}_m$ is the magnon spin current. Notice that Eq. (1) describes the time dependent behavior of the magnetization whereas the relatively long range magnetization correlation describes the SW propagation. Though the z-component of the magnetization across the Néel wall is antisymmetric with respect to the wall center and can be described by $m_z(x) \propto -\arctan(x)$, the propagating SW has decaying amplitude thus the magnonic spin torque breaks the antisymmetric structure of the DW. Fig. 3 shows the simulated magnetization configurations of the NiFe in the case Fig. 2(d) with $f$ = 20 GHz, $H_0$ = 288.8 mT. The localized oscillation motions of the DW induced by the magnonic STT of SW and the torque of effective field are accompanied by in-plane counterclockwise rotation of the magnetization inside DW. Once rotated by 90° and becomes a Block wall, the DW shows a backward motion towards the SW source. We find that the propagating SW can drive a DW motion depending on the in-plane rotation of the magnetization at the center of the DW. Moreover, when SW propagates across the DW, we find the transmitted amplitude of SW is determined by the orientation of the DW magnetization. As presented in the supplemental material, we have formulated an equation to calculate the rotation at the center of the DW plane from the simulation results and performed self-consistency check about the DW velocities.

The impact of the DW orientation on the SW transmission is presented in the following. A spin wave is excited above the Walker breakdown threshold as illustrated in Fig. 2(c). Upon the incidence of the SW, DW starts to move and the structure is modified. We focus on the SW dynamics between $x$ = -0.5 to 0.5 µm here. We find the DW transmission ratio, $T_{DW}$, of the SW passing through the DW is a function of $\delta\varphi$. Note the $T_{DW}$ we concern here is the SW amplitude ratio with and without DW on the +x side. It is not the SW amplitude ratio after and before the DW. The first panel in Fig. 4(a) shows the spatial variation of the normalized $M_y$ component without any DWs while the SW propagates. The SW amplitude decays exponentially away from the source due to the damping in the LLG equation. The instantaneous magnetization of the small time varying component is given by $\boldsymbol{m} = \boldsymbol{m}_0 \exp[-i(kx - \omega t)]\exp(-x/\Lambda)$ where $m_0$ is the SW amplitude at the source, and $\Lambda$ is the attenuation length. In a uniform single domain, the attenuation length depends on α, an intrinsic parameter of the material and the dispersion relation. In the presence of a DW, the effective anisotropy field exerts a torque to change $\delta\varphi$ when the SW travels in the DW, giving rise to a demagnetization field inside the DW region as $\vec{H}_d = -M_S(N_x cos\delta\varphi \hat{x} + N_y sin\delta\varphi \hat{y})$. The demagnetization factor $N_x$ and $N_y$ are determined by the DW width and the line width, respectively. For very narrow wires, $N_y$ may be close to unity. Taking this field into account, the attenuation length depends on the change of $\delta\varphi$. As we show in the supplemental material [31], the $T_{DW}$, defined as the spin wave amplitude ratio with and without DW on the +x side of the DW, can be written as

$$T_{\mathrm{DW}} = e^{-2\Delta \left| \frac{1}{\Lambda_D} - \frac{1}{\Lambda_0} \right|}, \tag{2}$$

where $\Delta$ is the width of the DW, $\Lambda_D$ is the attenuation length inside the DW, and $\Lambda_0$ is the attenuation length outside the DW. When the SW propagates across the DW, we find $T_{DW}$ is a function of $\delta\varphi$. Previous studies compared the SW wavelength and the DW width and found high $T_{DW}$ for small SW wavelengths [12, 13]. We found that the DW orientation is a decisive factor especially at low SW frequencies. The calculated $T_{DW}$ is plotted in Fig. 4(b) as blue circles. There are two minima at $\delta\varphi = 45°$, 225° and high, oscillatory values in the second and fourth quadrants. Here we treat the DW rigidly with one single $\delta\varphi$. Variation of $\delta\varphi$ inside a DW will make the analysis much more complicated.

The simulation results for NiFe indeed show more interesting behavior. Here $T_{DW}$ is defined as the SW amplitude at the position $x = 0.5$ µm divided by the amplitude without a DW, regardless the DW position. Fig. 4(a) shows the snapshots of the SW amplitude (normalized $m_y$ component) in the nanostrips without and with a DW with selected values of $\delta\varphi$, 45° and 140°. The shaded areas indicate the position of the DWs. For the 45° DW $T_{DW} = 0.016$, the SW is almost unable to propagate across the DW and most of the SW is reflected. However, for the 140° DW, the SW can penetrate the DW with relatively high transmission, $T_{DW} = 0.822$. The red triangle symbols in Fig. 4(b) show the simulation results of DW angular dependence of $T_{DW}$ for a propagating SW with $f = 20$ GHz and $H_0 = 288.8$ mT. It is strongly anisotropic. The minimum $T_{DW}$ occurs at -10°, 45°, 175°, and 225° when the DW speed or acceleration is maximum [31]. For CoFeB, though we did not find localized oscillatory motion with $f = 5$ GHz, our simulations still show very similar angular dependence on DW oscillation behavior with minimum $T_{DW} = 0.13$ when $\delta\varphi = 45°$ and maximum $T_{DW} = 0.98$ when $\delta\varphi = 135°$. The DW serves as SW switch at relatively low frequencies and results in rich DW dynamics. Thus, $\delta\varphi$ acts as an amplitude filter for SW propagation in the PMA nanostrips and can be engineered to control SWs in practical devices.

We presented two PMA materials. CoFeB is a popular room temperature PMA material widely used in academic research and in industry. NiFe was reported to show PMA only at low temperature [29]. Both materials show similar response with different characteristic SW frequencies and amplitudes. The DW motion and the SW transmission ratio through DW can be utilized in spintronic devices. Should the DW motion be the desired property, large SW amplitude in the range $H_0 = 100$ mT would be required. This is available with the advent of rf technology. The variation of SW transmission ratio through DW is readily applicable for wide ranges of SW frequency and amplitude. Many parameters can be fine-tuned for applications. For example, materials parameters like the exchange stiffness constant $A$, the perpendicular anisotropy constant $K_\perp$, nanostrip dimension and the configuration of SW source and DW, etc. Optimization for each specific application is feasible.

In summary, we find the SW influence at a DW induces an effective field torque, which leads to the rotation of the domain wall plane. The forward DW motion is a contribution of the demagnetization field due to the increase of transverse components of magnetization in the DW region, and the $V_{DW}$ is dependent on $\delta\varphi$. The transmission ratios of the SW are determined by the magnetization orientation of the DW and show complicated dependence at low frequencies. We can thus manipulate the DW motion by selecting the SW frequency and/or controlling the SW amplitude by designing a DW angle. The interplay between a SW and the DW offers rich dimensions for circuit design in spintronics.

Acknowledgment:

This work was financially supported by the Ministry of Science and Technology NSC 102-2112-M-001-021-MY3 and the Academia Sinica, Taiwan.

Figure Captions

Fig. 1. (color online) The sample geometry for simulation is a nanostrip 4 µm long, with a Néel domain wall located at the center, and a spin wave source 0.5 µm to the left. The cones indicate the precession of the magnetizations. $\delta\varphi$ is the rotation of the magnetization at the center of the domain wall. $\boldsymbol{\tau}_d$, $\boldsymbol{h}_d$, $\boldsymbol{\tau}_K$, and $\boldsymbol{h}_K$ are the torques and effective fields due to demagnetization and anisotropy, respectively. Drawing not to scale.

Fig. 2. (color online) (a) Typical domain wall displacement as functions of simulation time and magnetic field amplitude of the spin wave with frequencies of 5 or 6 GHz for CoFeB. (b) to (e), selected fields 200 mT, 280 mT, 288.8 mT, and 320 mT, with 20 GHz SW showing different types of domain wall motion for NiFe. (f) phase diagram of the domain wall displacement as functions of SW amplitudes and time.

Fig. 3. (color online) Top view of the configuration of the y component of the magnetization and the DW motion in the case of Fig. 2(d). On the left is the simulation time of the snapshot. The black arrow indicate the orientation at the center of the DW, also noted on the right.

Fig. 4. (color online) Transmission of SW in the case of Fig. 2(d). (a) Spatial variation of the normalized $m_y$ component without and with DW for $\delta\varphi = 45°$ and 140°. Shaded areas indicate the domain wall region. (b) Polar plot of the transmission ratio versus $\delta\varphi$. Blue circles are calculated results and the red triangles are from simulations.

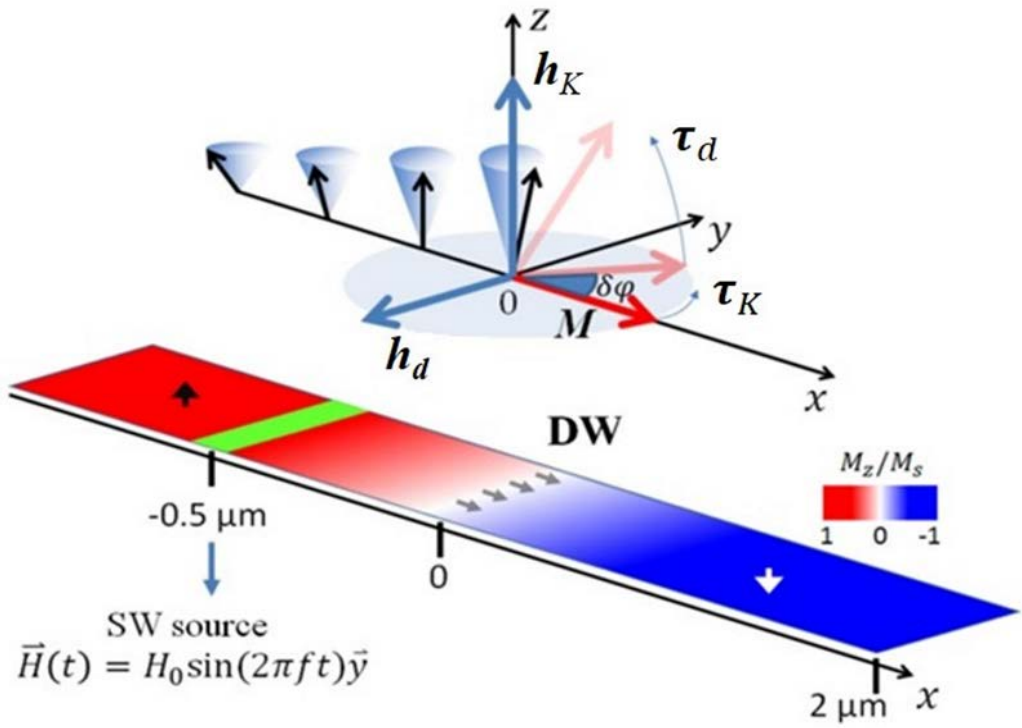

Fig. 1

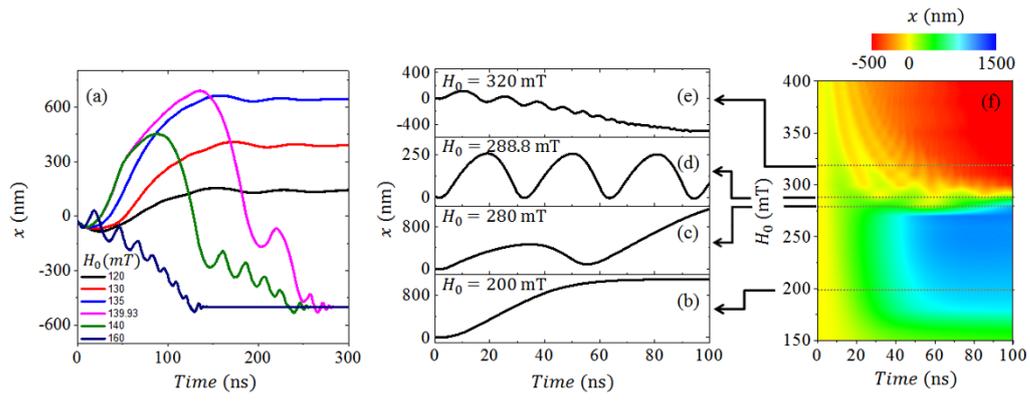

Fig. 2

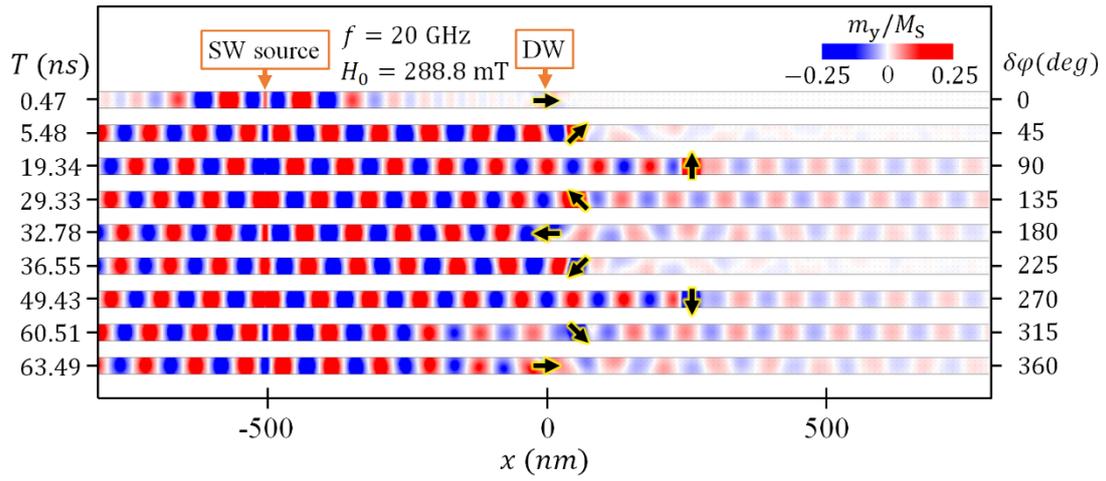

Fig. 3

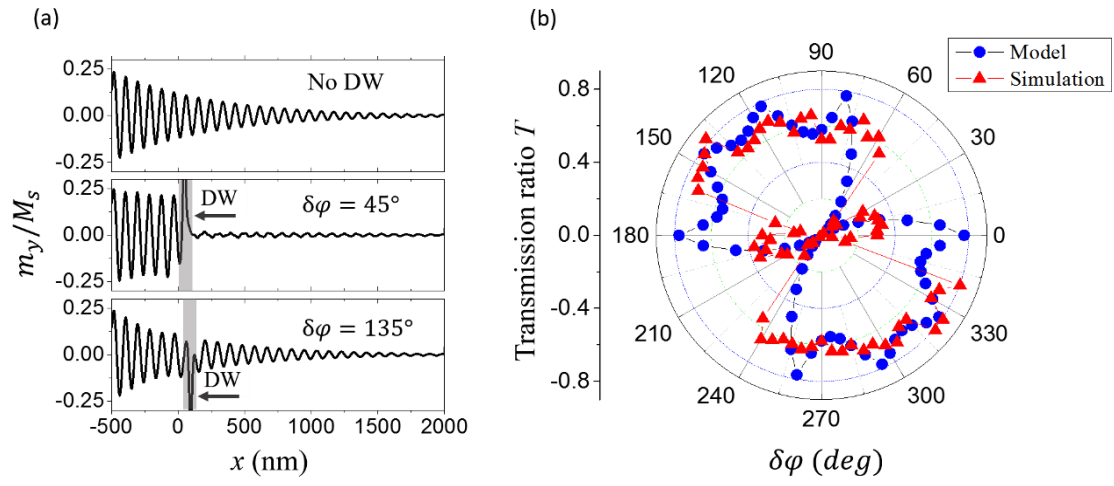

Fig. 4